\documentclass[%
 reprint,
 amsmath,amssymb,
 aps,
 pre,
 showpacs
]{revtex4-1}

\usepackage{graphicx}% Include figure files
\usepackage{dcolumn}% Align table columns on decimal point
\usepackage{bm}% bold math
\usepackage{caption}
\usepackage{subcaption}
\usepackage{parskip}

\begin{document}

\title{Spectral partitioning in equitable graphs}
\author{Paolo Barucca}
\affiliation{University of Zurich and London Institute for Mathematical Sciences}

\date{\today} 

\begin{abstract}
Graph partitioning problems emerge in a wide variety of complex systems, ranging from biology to finance, but can be rigorously analyzed and solved only for a few graph ensembles. Here, an ensemble of equitable graphs, i.e. random graphs with a block-regular structure, is studied, for which analytical results can be obtained.
In particular, the spectral density of this ensemble is computed exactly for a modular and bipartite structure. Kesten-McKay's law for random regular graphs is found analytically to apply also for modular and bipartite structures when blocks are homogeneous. Exact solution to graph partitioning for two equal-sized communities is proposed and verified numerically, and a conjecture on the absence of an efficient recovery detectability transition in equitable graphs is suggested. 
Final discussion summarizes results and outlines their relevance for the solution of graph partitioning problems in other graph ensembles, in particular for the study of detectability thresholds and resolution limits in stochastic block models.
\end{abstract}

\maketitle

\section{\label{sec:intro}Introduction}

The recent developments of network theory driven by the increasing number of applications in biology, ecology, social systems, economics and finance \cite{newman2006structure,easley2010networks}, have stimulated theoretical research in graph theory. In particular, the need to establish the statistical significance of various network metrics and properties in real systems has ignited new results in statistical inference \cite{decelle2011inference}, spectral theory of random graphs \cite{rogers2008cavity,chung2003eigenvalues,erdos2013spectral,farkas2001spectra,bordenave2010resolvent,nadakuditi2012graph}, ensembles of exponential random graphs \cite{robins2007introduction}. \\
In most networks elements are divided into separate groups, and their behavior will often depend on this division. Finding an optimal partition then allows to better understand the mesoscopic dynamics of the system and to obtain a more efficient reduced representation in terms of interacting groups. As a consequence, community detection has become a pivotal topic in network science. Stochastic block models (SBM) \cite{holland1983stochastic,karrer2011stochastic} have been introduced to understand, model, and analyze communities, and have allowed to gain theoretical insight on the performance and limitations of graph partitioning algorithms.
Recently it has been shown that recovery of communities in SBM displays a detectability transition in the sparse regime, i.e. when the edges are few, communities can be too weak to be identifiable \cite{decelle2011inference,decelle2011asymptotic}.
Equitable graphs, the family of graph ensembles analyzed in this work, represents a block-regular counterpart of the long-studied SBM: in SBM edges are drawn independently with a probability depending on the assignment of the two terminal nodes, which results in a Poisson distribution of the number of edges between a pair of groups; in equitable graphs the number of intra-block and inter-blocks edges are fixed for each node and the graph is the result of a random matching between such edges. This class of random graph models has been first analyzed in \cite{radicchi2013detectability} in a dense approximation, in \cite{newman2014equitable} for the first time under the name of equitable graphs and under the name microcanonical stochastic block model in \cite{peixoto2016nonparametric}. In \cite{brito2016recovery} it has been shown that equitable graphs with two-equally-sized communities have a unique partition almost surely in the large size limit, that an efficient algorithm can be derived in a large region of the ensemble's parameters, and that full recovery of communities can be obtained starting from a group assignment with an extensive overlap with the original partition.
In this paper, spectral theory of random graphs is used to disentangle noise and signal in equitable graphs, and an efficient algorithm for full recovery of communities is proposed. Following the derivation in \cite{rogers2008cavity} a finite set of non-linear equations for the spectral density is obtained, analogously to \cite{newman2014equitable}. The solution is found to obey the expected Kesten-McKay's law for regular graphs, allowing to analytically predict the failure of naive spectral partitioning based on the second eigenvector of the adjacency matrix.\\
The paper is organized as follows: in Section \ref{sec:reg} equitable graphs are defined and the inference problem is introduced. In Section \ref{sec:spec} a brief introduction of the cavity approach to the spectral density is provided and the general expression for the cavity variances of equitable graphs is derived. In Section \ref{sec:res} the expression is solved for modular and bipartite two-community structures and its consequences on spectral clustering are outlined. A general methodology for the block structure inference in equitable graphs based on eigenvectors's extendedness is introduced to overcome the limitation of naive spectral partitioning. Numerical evidence of all results is presented. \\ Finally in Section \ref{sec:con} the relevance of the results with respect to open questions in theory of random graphs is discussed, and possible directions of research both for analytical results in spectral theory and for spectral methodologies for graph partitioning are outlined.

\section{\label{sec:reg}Equitable random graphs}

\begin{figure}\label{fig:adj}

   \begin{subfigure}[b]{0.5\textwidth} 
   \includegraphics[width=1\linewidth]{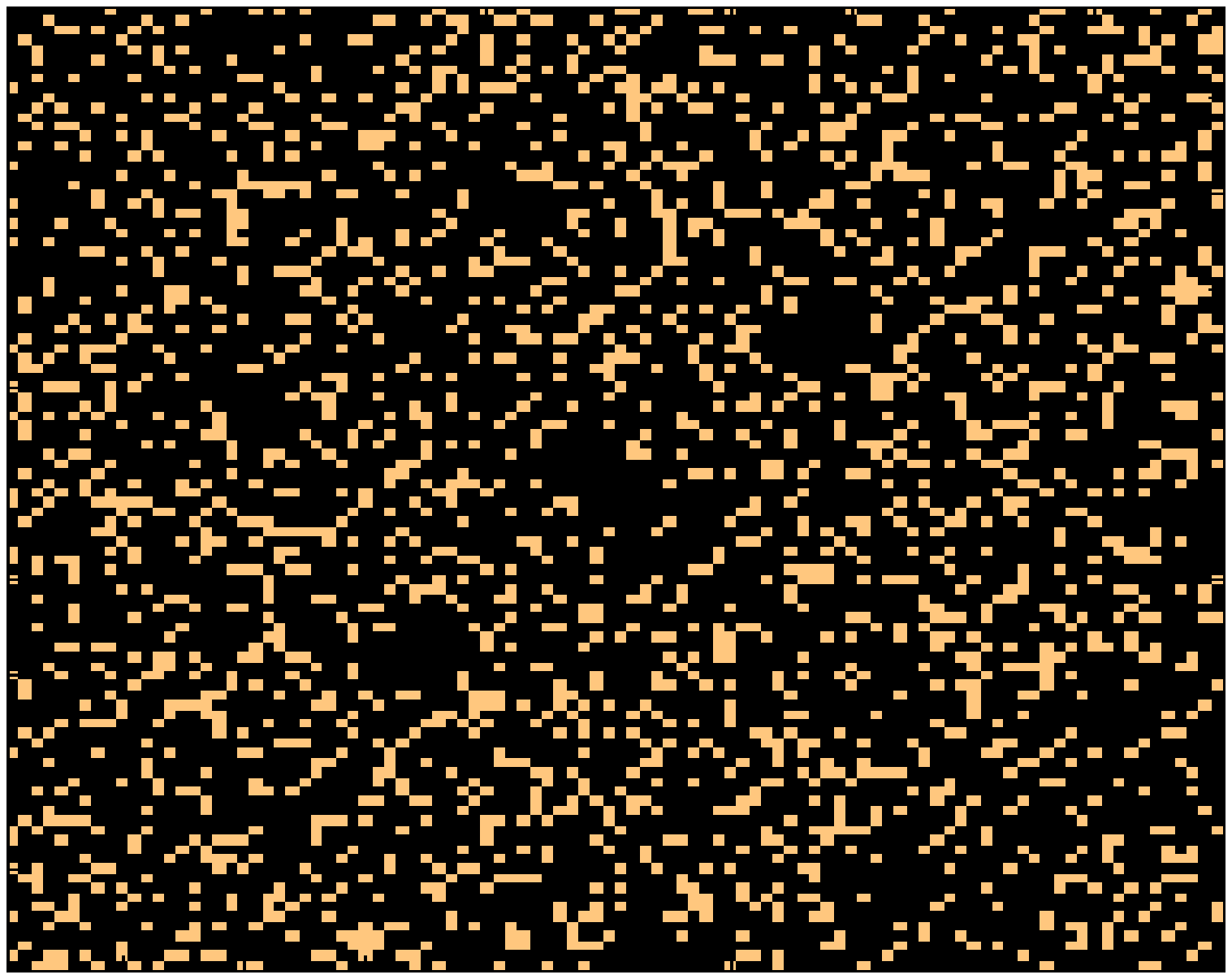}
   \caption{\centering}
   \label{fig:Ng1} 
\end{subfigure}
\begin{subfigure}[b]{0.5\textwidth}
   \includegraphics[width=1\linewidth]{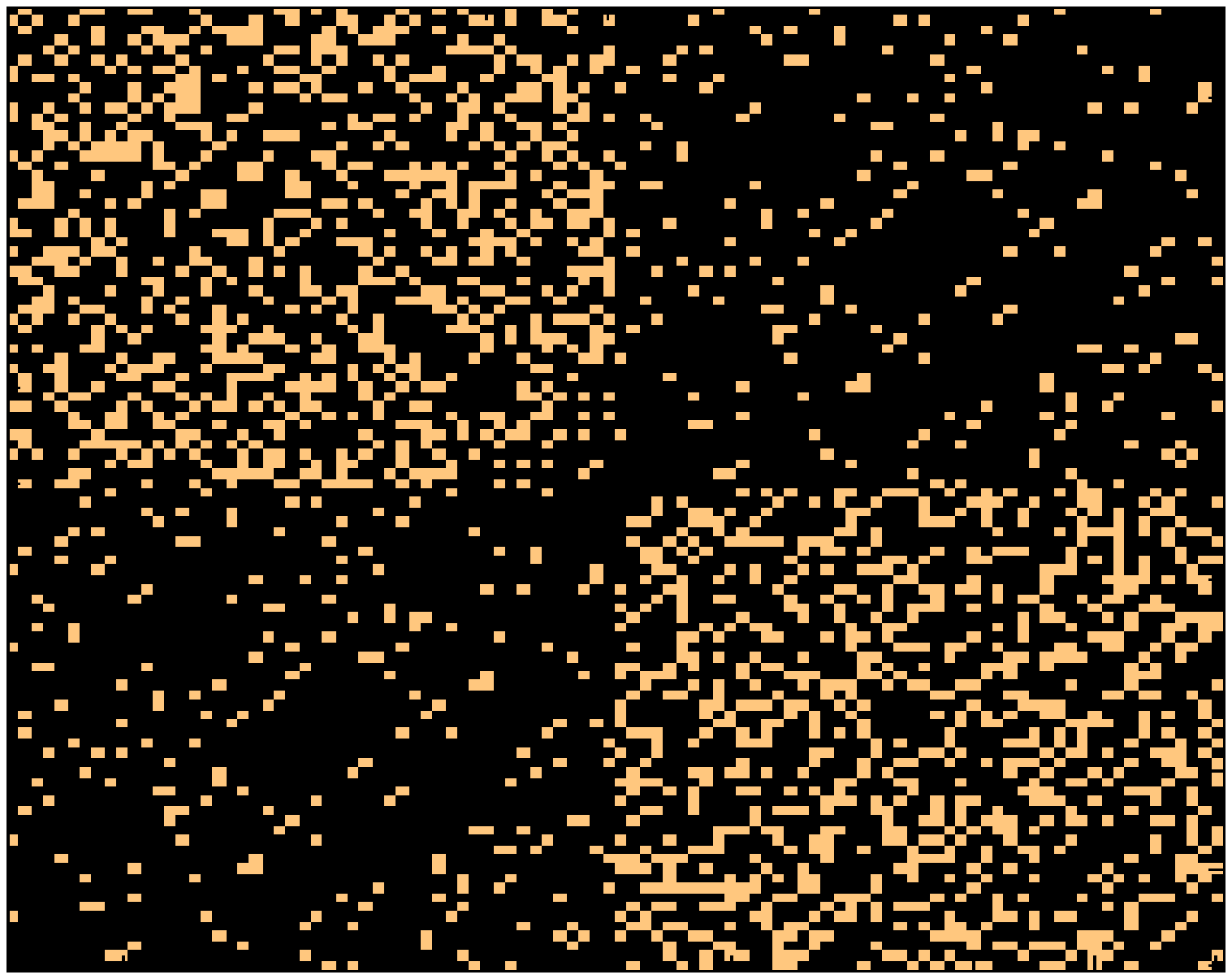}
   \caption{\centering}
   \label{fig:Ng2}
\end{subfigure}

\caption[Inference]{(a) Adjacency matrix of a random regular block model graph with a community structure where no block structure seems to be present even though the connectivity matrix reads, $\mathbf{c} = [16,\; 4;\; 4,\, 16]$. \\(b) Same graph but rows and columns of the adjacency matrix are sorted according to the block structure, which becomes evident.}
\end{figure}

An ensemble of equitable graphs is defined by a set of vertices $V$, a partition $\mathbf{B}=\{B_a\}_{a=1}^m$ dividing $V$ in $m$ non-overlapping sets of vertices, also called blocks, and a connectivity matrix $\mathbf{c}$, a $m\times m$ matrix of non-negative integer numbers \cite{peixoto2016nonparametric}. For the sake of simplicity in the following I will refer to block $B_a$ with its corresponding integer index $a$. For later use, I also introduce for all nodes the assignments $g_i$, such that for each node $i$ holds $i\in B_{g_i}$. \\ 
Each graph $G=(V,\,E)$ of a random regular block model must satisfy the constraints:

\begin{equation}\label{eq:equitabledef}
\forall B_a,B_b \in \mathbf{B}\:\forall i \in B_a \, \, \,  |\{(i,j) \in E\,|\,j \in B_b \}| = c_{ab},
\end{equation}

i.e. the total number of edges of node $i$ in block $B_a$ with a vertex in $B_b$ equals $c_{ab}$, for every vertex $i$ and every pair of blocks $B_a$ and $B_b$. Eq.\eqref{eq:equitabledef} means that all nodes in a given block share the same connection pattern, i.e. number of links with each other block. This condition is stronger than a simple regularity within blocks, i.e. where all nodes in a block only share the same total number of links. 
In the case of blocks of different sizes, $|B_a|=N_a$ such that $\sum_{a=1}^mN_a=|V|$, then, for the system to have solution the connectivity matrix $\mathbf{c}$ and block sizes must obey the relations

\begin{equation}\label{eq:equitablecon}
\forall B_a,B_b \in \mathbf{B}\:N_ac_{ab} = N_bc_{ba},
\end{equation}

i.e. the total number of edges between blocks $a$ and $b$ must be uniquely defined. \\
All graphs satisfying (\ref{eq:equitabledef}) have equal probability in the ensemble. \\
If I introduce the block degrees $k_{i\rightarrow a} =  |\{(i,j) \in E | j \in B_a \}| $, (\ref{eq:equitabledef}) can be reformulated as follows: the vector of block degrees of each node in a given block equals the row of the connectivity matrix corresponding to the block index, i.e. $\forall i \in B_a\:k_{i\rightarrow B_b} = c_{ab}$.  \\
Both stochastic block models and equitable graphs are based on an analogous set of parameters, i.e. block assignments and connectivity matrix, nevertheless there is no trivial mapping between stochastic block models, which are defined via link probabilities and their regular counterpart, which are defined via (\ref{eq:equitablecon}).
A useful analogy to make sense of their relationship is the following: stochastic block models \cite{holland1983stochastic} correspond to random regular block models as the Erdos-Renyi random graphs correspond to the k-regular random graphs, in the sense that in both cases the randomness which is eliminated from the ensemble is the one given by the (block) degree distribution. \\
A different ensemble of random graphs with a block structure is the regular stochastic block models, studied in \cite{decelle2011inference, kawamoto2015limitations}, where the probability measure is the same as in stochastic block models but a regularity constraint is imposed to all nodes.\\ 
Moreover, the form of the constraints in (\ref{eq:equitablecon}) allow edges to be drawn independently for each pair of blocks, and, for the case of blocks of the same size, it is possible to sample equitable graphs simply by assembling regular graphs: between each pair of blocks the edges are drawn according to a k-regular graph, where the value of $k$ equals the corresponding element of the connectivity matrix, then the total set of edges is given by the union of the sets of edges for each of the $m$ regular graphs and $m*(m-1)$ bi-regular graphs. \\
In the latter the focus will be entirely on the representation of $G$ in terms of its adjacency matrix $A=(a_{ij})_{i,j=1}^N$ where,
\[
    a_{ij} = \begin{cases}
        1, & \text{if}\:\:(i,j)\in E\\
        0, & \text{otherwise}
        \end{cases}
  \]
which allows to compute graph properties in algebraic form and can also be used to visualize the inference problem associated with the graph (Figs.\ref{fig:Ng1}-\ref{fig:Ng2}): when parameters are unknown, there is no a-priori criterion to sort indices and the non-zero elements of the adjacency matrix do not display any specific block structure; once the parameters are known, rows and columns can be sorted according to the block indices, and the structure arises in a clear manner.\\

\subsection*{The inference problem}
Given an equitable graph $G$, the inference problem consists in reconstructing the parameters, i.e. the partition $B$ and the connectivity matrix $\mathbf{c}$, that generated the graph.  \\
I study inference on this ensemble of random graphs because it allows to analyze the performance of different algorithms in absence of the noise coming from degree heterogeneity. In SBM there exists a sharp transition in the assortativity parameter, first conjectured in \cite{decelle2011asymptotic} and later proved rigorously in a series of works that demonstrate both that asymptotically (i) below such threshold, recovery is information theoretically impossible \cite{mossel2015reconstruction} while (ii) above, an efficient algorithm finds a partition with positive overlap with the original one \cite{mossel2013proof,massoulie2014community}. In this work, it is shown that the regularity condition, as also found in \cite{brito2016recovery}, substantially change the detectability properties of the ensemble.

\section{\label{sec:spec}Spectral theory}

In this section, the spectral properties of random regular block model graphs with blocks of same size are investigated. Analytical results are presented, both for the discrete and the continuous part of the spectrum. 
\subsection{Discrete part: the signal}
Here is shown how isolated eigenvectors of the adjacency matrix entail exact informations on the block structure of equitable graphs. Starting from the secular equation
\begin{equation}\label{eq:secEqA}
\underset{j=1}{\overset{N}{\sum}}a_{ij}u_{j}=\lambda u_{i},
\end{equation}
an ansatz of \textit{block-symmetry} can be made such that nodes in the same block share the same eigencomponent, i.e. for all $i$ is hypothesized that $u_i = u_{g_i}$. Since the number of neighbors
between different groups is fixed it follows that:
\begin{equation}\label{eq:secEqC}
\underset{b=1}{\overset{m}{\sum}}c_{ab}u_{b}=\lambda u_{a},
\end{equation}
which yields the useful conclusion, also pointed out in \cite{brito2016recovery,newman2014equitable} that each block-symmetric eigenvector of the adjacency matrix corresponds to an eigenvector of the connectivity matrix $\mathbf{c}$, and viceversa. These eigenvectors correspond to a finite set of non-densely distributed, at most finitely degenerate eigenvalues. Generally, they can be positioned everywhere in the spectrum and when the block structure is particularly weak they will lie within the bulk of the spectrum. I will refer to them as the community eigenvectors.

\subsection{Continuous part: the noise}

In this paragraph statistical physics techniques are used to compute the bulk of the spectrum of the adjacency matrices of equitable graphs. The derivation is entirely equivalent to the one already found in \cite{newman2014equitable}, here I simply report it in terms of cavity variances \cite{rogers2008cavity}, for the reader's convenience.
Given an ensemble of $N\times N$ symmetric matrices the set of eigenvalues of a given adjacency matrix $A$ is denoted by $\{\lambda_i^A\}_{i=1}^N$. The corresponding empirical spectral density is defined as:
\begin{equation}\label{eq:specDen}
\rho(\lambda;A)=\frac{1}{N}\sum_{i=1}^N\delta(\lambda-\lambda_i^A),
\end{equation}
which satisfies the identity \cite{edwards1976eigenvalue}:
\begin{equation}\label{eq:edwards}
\rho(\lambda;A)=\frac{2}{\pi}\lim_{\epsilon\rightarrow 0^+}\frac{1}{N}\Im\left[\frac{\partial}{\partial z}\log\mathcal{Z}(z;A)\right]_{z=\lambda-i\epsilon}
\end{equation}
where $\Im[z]$ denotes the imaginary part of $z$ and where $\log\mathcal{Z}(z;A)$ is obtained via Gaussian integrals as in\cite{edwards1976eigenvalue}, i.e.: 
\begin{equation}\label{eq:Z}
\mathcal{Z}(z;A)=\int \left[\Pi_{i=1}^N\frac{dx_i}{\sqrt(2\pi)} e^{-H(x;z,A)}\right]
\end{equation}
with $H(x;z,A) = \frac{z}{2}\sum_{i}^N x_i^2  -\frac{1}{2}\sum_{i,j}^N A_{ij}x_ix_j $. Such formulation yields an expression for the spectral density of any graph of the ensemble in terms of the variances of the Gaussian variables introduced in (\ref{eq:Z}), 
\begin{equation}\label{eq:specDensVar}
\rho(\lambda;A)=\frac{1}{\pi}\lim_{\epsilon\rightarrow 0^+}\frac{1}{N}\Im\left[\sum_i^N\langle x_i^2\rangle_z\right]_{z=\lambda-i\epsilon}.
\end{equation}
In principle, computing variances in (\ref{eq:specDensVar}) is not easier than diagonalizing the adjacency matrix but for sparse graphs an approximation method has been proposed that holds exactly in the large $N$ limit, the cavity method \cite{mezard2002analytic,bordenave2010resolvent}.\\
In the cavity method, conditional probability distributions are introduced for each node and are parametrized by specific variables, i.e. the cavity variances $\Delta_i^{(j)}$, each representing the variance of $x_i$ if its neighbor $j$ is not taken into account.
With such approximation the following set of self-consistent equations can be derived \cite{rogers2008cavity}:
\begin{equation}\label{eq:cavityVar}
\Delta_{i}^{(j)}(z)=\frac{1}{z-\underset{l\in\partial i\setminus j}{\overset{N}{\sum}}A_{il}^{2}\Delta_{l}^{(i)}(z)},
\end{equation}
where $\partial i$ is the set of neighbor of node $i$, i.e. $\partial i = \{e \in E | i \in e\}$.
From cavity variances it is possible to compute node variances via the equations
\begin{equation}\label{eq:nodeVar}
\Delta_{i}(z)=\frac{1}{z-\underset{l\in\partial i}{\overset{N}{\sum}}A_{il}^{2}\Delta_{l}^{(i)}(z)},
\end{equation}
which lead to compute the spectral density $\rho(\lambda;A)$. \\
In the case of equitable graphs the ansatz of \textit{block-symmetry} can be made for the cavity variances:
\begin{equation}\label{eq:ansatz}
\Delta_{i}^{(j)}(z) = \Delta_{g_i}^{(g_j)}(z).
\end{equation}
This ansatz, also made in \cite{newman2014equitable}, allows to perform the summation in the denominator, that consistently turns out to be independent from the individual node, but only from its block, thus reducing the set of equations for cavity variances from a size of order $N$ (in the sparse case) to the following set of $m^2$ equations:
\begin{equation}\label{eq:ansatzCav}
\Delta_{a}^{(b)}(z)=\frac{1}{z-\underset{c}{\overset{m}{\sum}}(c_{ac}-\delta_{bc})^+\Delta_{c}^{(a)}(z)},
\end{equation}
where $(x)^+ = \max(x,0)$.
Block variances can then be computed,
\begin{equation}\label{eq:blockVar}
\Delta_{a}(z)=\frac{1}{z-\underset{c}{\overset{m}{\sum}}c_{ac}\Delta_{c}^{(a)}(z)},
\end{equation}
and finally the spectral density,
\begin{equation}\label{eq:equitablespec}
\rho(\lambda)=\frac{1}{\pi m}\underset{a=1}{\overset{m}{\sum}}\Im[\Delta_{a}(z)]_{z=\lambda-i\epsilon}.
\end{equation}

\section{\label{sec:res}Results}

In this section I derive the threshold at which naive spectral partitioning fails and introduce a general algorithm based on the inverse participation ratio (IPR) to solve the inference problem in equitable graphs with two communities. \\

\subsection{\label{sec:res_mod}Modular structures}

\begin{figure}
\includegraphics[width=80mm]{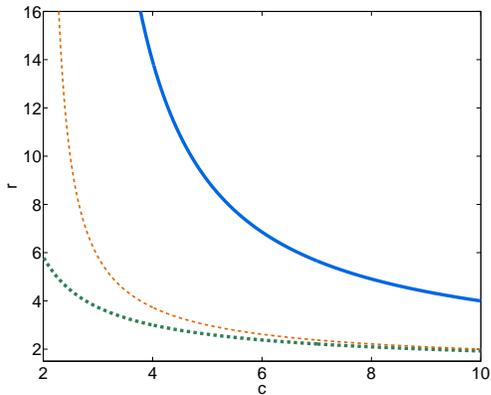} \protect\caption{Critical lines in the plane $(c,\,r)$ for equitable graphs (solid line), two-block random regular graphs (dashed line), and SBM (squared line). Above the solid line standard spectral bisection works for equitable graphs. Below the solid line naive spectral bisection fails but the IPR-based algorithm succeeds in full recovery.}
\label{fig:critical} 
\end{figure}

\begin{figure}
\includegraphics[width=80mm]{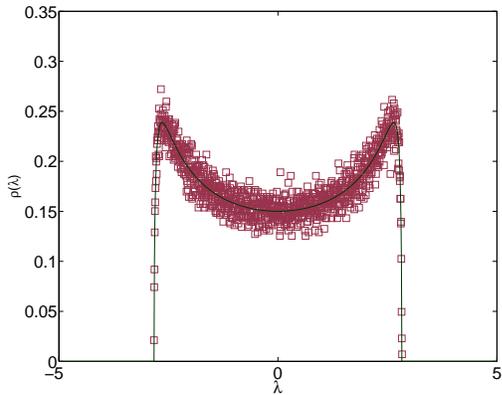} \protect\caption{(Modular case). Spectral density for $c=3$ and $r=2$, it corresponds to Kesten-McKay's law for a k-regular graph with $k=3$. Squares come from numerical diagonalization of a sample of $100$ equitable graphs of size $N=1000$.}
\label{fig:modDOE} 
\end{figure}

Graph partitioning, and in particular spectral bisection, is a long-standing problem in graph theory \cite{fortunato2010community,boppana1987eigenvalues,pothen1990partitioning,bollobas1999exact}.
Here for the modular case, two homogenous blocks are analyzed: the blocks share the same size and the connectivity matrix reads:
\begin{equation}
\mathbf{c}=\left(\begin{array}{ccc}
c_{in} & c_{out}\\
c_{out} & c_{in}
\end{array}\right)\label{eq:affmat}
\end{equation}
where $c_{in}$ and $c_{out}$ are non-negative integers such that $c_{in}>c_{out}$. In this case, all nodes in this ensemble of equitable graphs share the same total degree, $c = c_{in}+c_{out}$, so that the ensemble is a subset of c-regular graphs, also called regular stochastic block model \cite{brito2016recovery}.
For later use, I also define $r=c_{in}/c_{out}$, $\epsilon = 1/r$ is a quantity that has been used to characterize the strength of the assortative structure \cite{decelle2011asymptotic}.\\
Assortative structures have been widely investigated with various approaches: spectral methods \cite{krzakala2013spectral}, modularity maximization \cite{newman2016community}, belief-propagation \cite{decelle2011asymptotic}, Markov-chain Monte Carlo methods \cite{peixoto2013parsimonious}, and other heuristic algorithms \cite{blondel2008fast}. Stochastic block models have been shown to display a detectability transition \cite{decelle2011asymptotic,mossel2013proof}. \\
In this homogenous case cavity equations can be further simplified: cavity variances associated to the two blocks can be assumed to be equal, $\Delta_1^{(b)}=\Delta_2^{(b)}$ for $b=1,\,2$, and given the form of the equations, by inspection, it is also possible to look for fully-symmetric solutions such that $\Delta_{a}^{(b)} = \Delta^{(cav)}$ for all $(a,b)$. This ansatz yields:
\begin{equation}\label{eq:symmAnsatz}
\Delta^{(cav)}(z)=\frac{1}{z-(c_{out}+c_{out}r-1)\Delta^{(cav)}},
\end{equation}
Now the equation is identical to the one derived in \cite{rogers2008cavity}, and analogously carrying out the algebra, the spectral density found is the Kesten-McKay's law \cite{McKay1981expected}, as found in \cite{newman2014equitable}:
\begin{equation}\label{eq:Kesten-McKay}
\rho(\lambda)= \frac{c\sqrt{4(c-1)-\lambda^2}}{2\pi(c^2-\lambda^2)}
\end{equation}
where $c=c_{out}(1+r)$ (Fig.\ref{fig:modDOE}). \\
(\ref{eq:Kesten-McKay}) yields the maximal eigenvalue in the bulk,\\ $\lambda^{+}_{b} = 2\sqrt{c_{in}+c_{out}-1}$.
Community eigenvalues can be easily computed via the characteristic polynomial:
\begin{equation}\label{eq:charPolyn}
(rc_{out}-\lambda)^2 - c_{out}^2 =0.
\end{equation}
The first eigenvalue $\lambda_{max}$ equals the total connectivity $c$ and its corresponding eigenvector is constant and uninformative. On the other hand, the second community eigenvalue, $\lambda_{com}=c_{in}-c_{out}$, is informative and its relationship with $\lambda_{b}^+$ is crucial for the inference problem: when $\lambda_{com}>\lambda_{b}^+$ it is simply the second largest eigenvalue and its corresponding eigenvector can be easily and fast computed, but when $\lambda_{com}<\lambda_{b}^+$ then it is no longer the second eigenvalue and its ranking becomes unknown. This is exactly what is found in the numerical simulation in \cite{radicchi2013detectability}, which are based on the modularity matrix, $Q$. \\
Such transition occurs when $\lambda_{com}=\lambda_{b}^+$, which corresponds to the critical line in the plane $c-r$ (Fig.\ref{fig:critical}):

\begin{equation}\label{eq:criticalLine}
r_c = \frac{c+2\sqrt{c-1}}{c-2\sqrt{c-1}}
\end{equation}

\begin{figure}
\includegraphics[width=75mm]{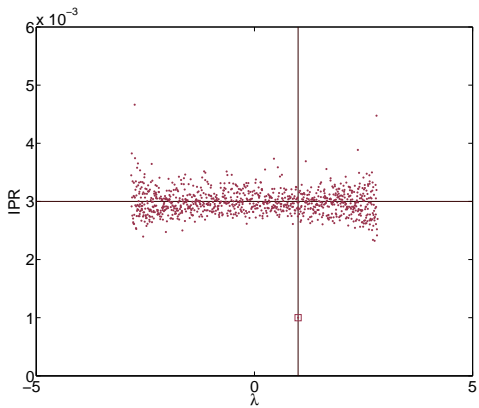} \protect\caption{(Modular case). Inverse participation ratio for each eigenvector in the plane $\lambda$-IPR. The eigenvectors of the bulk all share an IPR fluctuating around $3/N$ while the community eigenvector (square) has an IPR equal to $1/N$, which allows to solve the inference problem also when naive spectral partitioning fails. Parameters are $c=3$ and $r=2$.}
\label{fig:modIPR} 
\end{figure}

\begin{figure}
\includegraphics[width=80mm]{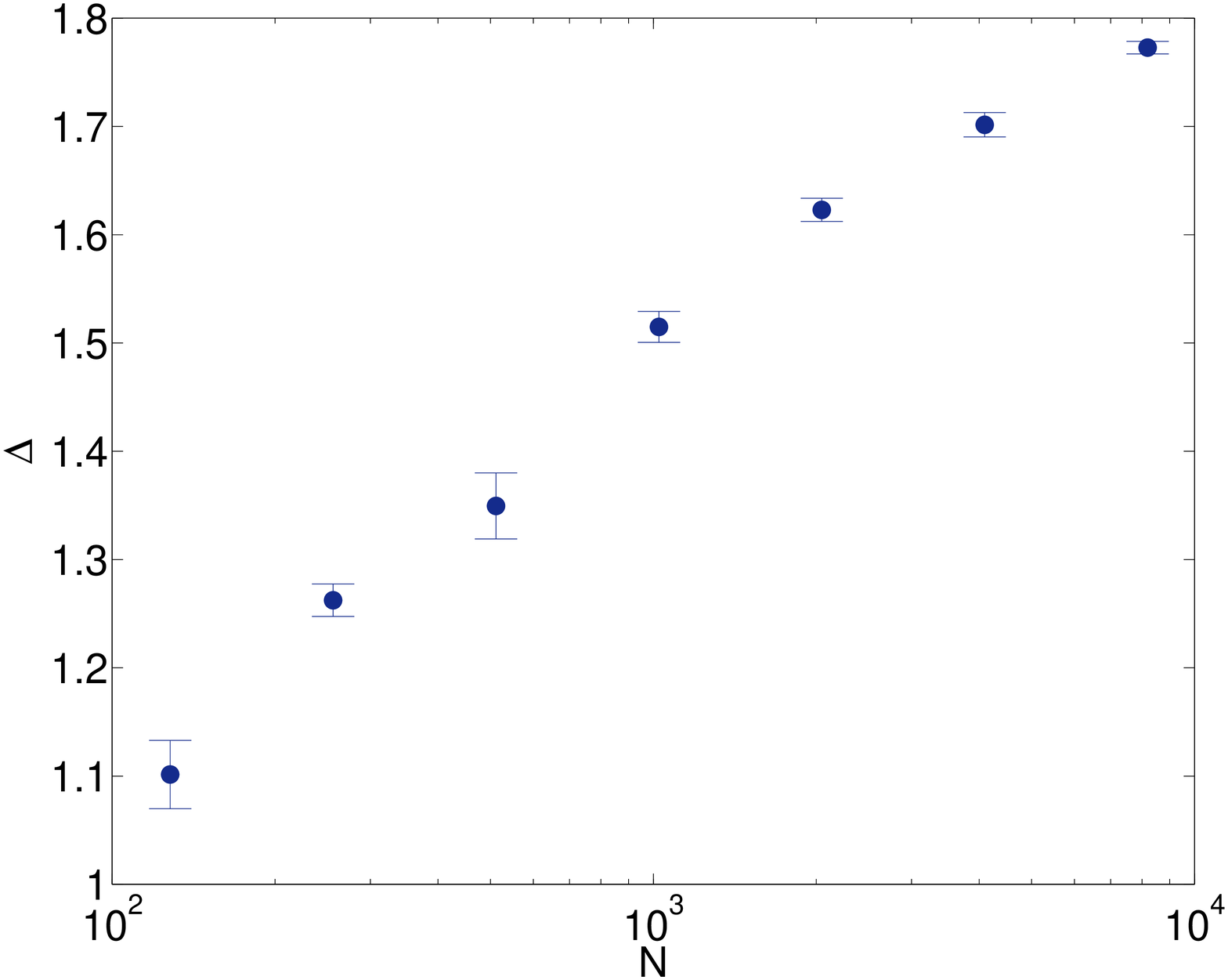} \protect\caption{(Modular case). Relative IPR divergence $\Delta$, for graphs of increasing size, ranging from $64$ to $8196$, each averaged over 10 samples. The most extended random eigenvector in the bulk always remains separate from the significant extended eigenvector associated to the community structure.
Parameters are $c=9$ and $r=2$. }
\label{fig:deltaScale} 
\end{figure}

Below the critical line, the community eigenvector gets lost in the bulk and a criterion is needed to identify the right eigenvector. \\
The solution to this detectability problem in equitable graphs can be found by exploiting the information about the eigenvectors. In fact, the distribution of the eigencomponents of the community eigenvector and of bulk's eigenvectors turn out to be significantly different: from the \textit{block-symmetry} ansatz, the eigenvector corresponding to $\lambda_{com}$ is more extended than the typical eigenvector of the bulk. \\
By looking at a measure of extendedness, such as the inverse participation ratio (IPR) \cite{coja2009spectral,kawamoto2015limitations}, it is possible to recognize the informative community eigenvector, $u_i^{com}$ associated to $\lambda_{com}$. In fact, the normalized community eigenvector is block-symmetric and all its elements scale like $1/\sqrt{N}$, i.e. $u_i^{com} = (\delta_{g_i1}-\delta_{g_i2})/\sqrt{N}$. Consequently,
\begin{equation}\label{eq:IPR}
IPR = \sum_{i}^N\left(u_i^{com}\right)^4 = \frac{N}{2}\sum_{a}^2\left(u_a\right)^4 = \frac{1}{N},
\end{equation}
independently from $r$. The inverse participation ratio of the community eigenvector is then $1/N$ while the random eigenvectors in the bulk have an expected IPR of $3/N$ \cite{dumitriu2012sparse} and a standard deviation of order $N^{-3/2}$, so that the signal-to-noise ratio grows with $N^{1/2}$. Therefore, for large graphs, the informative eigenvector remains distinguishable from a typical eigenvector from the bulk.\\
Then, the inference problem should be solvable for all values of $r>1$, i.e. as long as the signal is actually present, by searching for the most extended eigenvector of the adjacency matrix $A$ (Fig. \ref{fig:modIPR}), excluding the one associated with the uninformative maximum eigenvalue. An analogous approach was followed in \cite{coja2009spectral} to solve the planted coloring model, where color-symmetric eigenvectors were used to study the convergence of belief propagation in a special class of non-tree graphs, obeying a specific regularity condition. \\
This eigenvector-based solution for the inference problem would also solve the conundrum that arises in \cite{radicchi2013detectability}: the detectability threshold for the regular block model is found to be twice as large (see Fig.\ref{fig:critical}) as the one for the stochastic block model, even though the community structure in the former is partially deterministic \eqref{eq:equitabledef} while in the latter is entirely probabilistic. The detectability threshold in \cite{radicchi2013detectability} corresponding to \eqref{eq:criticalLine} only holds for spectral partitioning based on the second eigenvector of the adjacency matrix or, equivalently, on the first eigenvector of the modularity one. \\
This result is consistent with the one in \cite{brito2016recovery} stating that there exists a constant $d > 0$ such that, for $c_{in} > c_{out} > d $, the corresponding graph $G$ has a unique equitable partition asymptotically almost surely. Together, these two results constitute a solid basis for the general conjecture that equitable graphs with two equally-sized communities will always admit a full and efficient recovery of the original partition simply as soon as $c_{in} > c_{out} > 0$, whereas so far it has been rigorously proven only for a limited region of parameters \cite{brito2016recovery}. \\
To validate numerically the performance of the IPR-based algorithm on two-community equitable graphs, extensive simulations were performed varying the connectivity, the assortativity, and the size of the graph. In all cases considered, the algorithm yields an exact recovery, as long as $c_{in}>c_{out}$. To quantify the robustness of the result a measure of the distance between eigenvectors's statistics is introduced. IPRs of all eigenvectors (excluding the trivial constant eigenvector associated to the largest eigenvalues, $\lambda_1=c$) are ordered in increasing order, e.g. $IPR_2=1/N\leq IPR_3 \leq \dots $. Then, the relative IPR divergence, $\Delta$, is computed as the relative difference between the second smallest IPR in the sequence, constituting the most extended random eigenvector of the bulk and the minimum IPR, associated to the eigenvector that gathers the information on communities, i.e. $\Delta = (IPR_3 - IPR_2)/IPR_2$. In Fig.\ref{fig:deltaScale} it is shown how $\Delta$ grows with $N$, as fluctuations decrease, and then remains finite for large $N$. Since the Gaussian eigenvectors of the bulk have an average IPR of $3/N$ and the minimum possible IPR is $1/N$, the relative IPR divergence is upper-bounded by 2 in the asymptotic limit.

\subsection{\label{sec:res_bip}Bipartite structures}

\begin{figure}
\includegraphics[width=80mm]{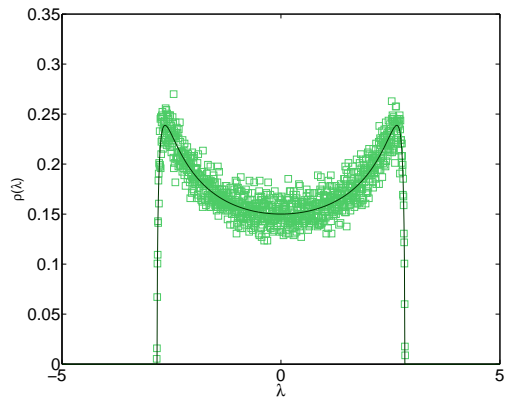} \protect\caption{(Bipartite case). Spectral density for $c_{in}=1$ and $c_{out}=2$, it corresponds to Kesten-McKay's law for a k-regular graph with $k=3$. Squares come from numerical diagonalization of a sample of $100$ equitable graphs of size $N=1000$.}
\label{fig:bipDOE} 
\end{figure}

In this paragraph I focus on disassortative equitable graphs, where edges within a block can be present but are always less than edges towards the other block, i.e. $c_{in}<c_{out}$.\\
The connectivity matrix $c_{ab}$ considered is the following:
\begin{equation}
\mathbf{c}=\left(\begin{array}{ccc}
c_{in} & c_{out}\\
c_{out} & c_{in}
\end{array}\right)\label{eq:affmat}
\end{equation}
where $c_{in}$ and $c_{out}$ are non-negative integers such that $c_{out}>c_{in}$. 
The analysis is entirely analogous to the one put forward for the assortative case, once the appropriate parallels are drawn: for large $c_{out}/c_{in}$ the informative community eigenvector corresponds to the lowest eigenvalue, $\lambda_{com} = c_{in}-c_{out}$, and the critical line is defined by the condition $\lambda_{com}=\lambda_b^-$. Also in this case the cavity equations admit a fully symmetric solution for the variances that leads to Kesten-McKay's law for the spectral density (Fig. \ref{fig:bipDOE}).

\begin{figure}
\includegraphics[width=75mm]{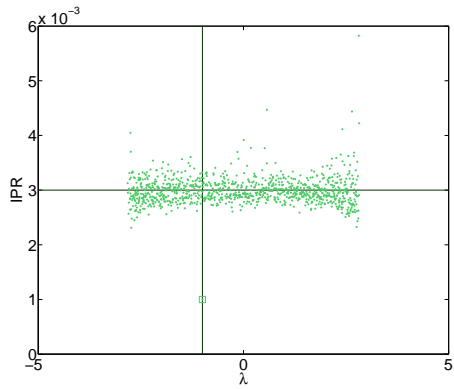} \protect\caption{(Bipartite case) Inverse participation ratio for each eigenvector in the plane $\lambda$-IPR. The eigenvectors of the bulk all share an IPR fluctuating around $3/N$ while the community eigenvector (square) has an IPR equal to $1/N$. Parameters are $c=3$ and $r=2$.}
\label{fig:bipIPR} 
\end{figure}

\section{\label{sec:con}Conclusions}

In this paper equitable graphs \cite{newman2014equitable} have been analyzed via spectral graph theory and graph partitioning theory. \\
In particular, in the framework of equitable graphs, the picture of the detectability threshold for naive spectral clustering, i.e. only using the eigenvector associated to the second largest eigenvalue of the adjacency matrix, emerges distinctly, as well as the crucial role of the statistics of eigenvectors. Strong analytical and numerical evidence has been provided in support of a new conjecture on the absence of an information-theoretic detectability transition in two-community equitable graphs.
Insights from equitable graphs could be used to develop new spectral methods based on both eigenvalues and eigenvectors properties in other graph ensembles.\\
Future work will deal with the interpolation between standard stochastic block models, regular stochastic block models, and equitable graphs. Further studies will be dedicated to the analysis of heterogeneous and multi-modular equitable graph, such as the equitable counterpart of planted partition model, in relation to the problem of resolution limit \cite{fortunato2007resolution} in modularity maximization, and to the generalization of the IPR based algorithm.

\section*{Acknowledgement}
The author acknowledges support from: FET Project DOLFINS nr. 640772, and FET IP Project MULTIPLEX nr. 317532.
The author would like to thank Ton Coolen, Travis Martin, Fabrizio Lillo, Elisa Letizia, Piero Mazzarisi and Daniele Tantari for fruitful discussions.

\end{document}